\documentclass[conference]{IEEEtran}
\IEEEoverridecommandlockouts
\newcommand{\tr}{\mathrm{tr}}

\usepackage{cite}
\usepackage{amsmath,amssymb,amsfonts}
\usepackage{algorithmic}
\usepackage{graphicx}
\usepackage{textcomp}
\usepackage{xcolor}
\usepackage{algorithm}
\usepackage{subcaption}
\usepackage{hyperref}
\usepackage{url}
\def\BibTeX{{\rm B\kern-.05em{\sc i\kern-.025em b}\kern-.08em
    T\kern-.1667em\lower.7ex\hbox{E}\kern-.125emX}}
\begin{document}

\title{QFlowNet: Fast, Diverse, and Efficient Unitary Synthesis with Generative Flow Networks}

\author{\IEEEauthorblockN{1\textsuperscript{st} Inhoe Koo}
\IEEEauthorblockA{\textit{NextQuantum and}\\ \textit{Department of Electrical and Computer Engineering}\\
\textit{Seoul National University}\\
Seoul, Korea\\
ihkoo711@snu.ac.kr}
\and
\IEEEauthorblockN{2\textsuperscript{nd} Hyunho Cha}
\IEEEauthorblockA{\textit{NextQuantum and}\\ \textit{Department of Electrical and Computer Engineering}\\
\textit{Seoul National University}\\
Seoul, Korea\\
ovalavo@snu.ac.kr}
\and
\IEEEauthorblockN{3\textsuperscript{rd} Jungwoo Lee}
\IEEEauthorblockA{\textit{NextQuantum and}\\ \textit{Department of Electrical and Computer Engineering}\\
\textit{Seoul National University}\\
Seoul, Korea\\
junglee@snu.ac.kr}
}
\author{
\IEEEauthorblockN{Inhoe Koo, Hyunho Cha, and Jungwoo Lee}
\IEEEauthorblockA{\textit{NextQuantum, Seoul National University, Republic of Korea}}
}
\maketitle

\begin{abstract}
Unitary Synthesis, the decomposition of a unitary matrix into a sequence of quantum gates, is a fundamental challenge in quantum compilation. Prevailing reinforcement learning (RL) approaches are often hampered by sparse reward signals, which necessitate complex reward shaping or long training times, and typically converge to a single policy, lacking solution diversity. In this work, we propose QFlowNet, a novel framework that learns efficiently from sparse signals by pairing a Generative Flow Network (GFlowNet) with Transformers. Our approach addresses two key challenges. First, the GFlowNet framework is fundamentally designed to learn a diverse policy that samples solutions proportional to their reward, overcoming the single-solution limitation of RL while offering faster inference than other generative models like diffusion. Second, the Transformers act as a powerful encoder, capturing the non-local structure of unitary matrices and compressing a high-dimensional state into a dense latent representation for the policy network. Our agent achieves an overall success rate of 99.7\% on a 3-qubit benchmark (lengths 1-12) and discovers a diverse set of compact circuits, establishing QFlowNet as an efficient and diverse paradigm for unitary synthesis.
\end{abstract}


\section{Introduction}
The translation of quantum algorithms into executable operations on physical hardware is a critical step enabled by quantum compilation \cite{b1}. A fundamental sub-task within this process is Unitary Synthesis, the decomposition of a given unitary matrix $U$ into a sequence of gates from a basis gate set. This task is broadly divided into two categories: \emph{exact synthesis}, which seeks a circuit $V_{C}$ that precisely matches the target $U$, and \emph{approximate synthesis}, which aims to find a circuit that is close enough within a given precision $\epsilon$ \cite{b2, b3}. Our work, similar to several other recent ML-based approaches \cite{b4, b5}, focuses on the challenge of exact synthesis. The primary difficulty of this task lies in its combinatorial nature. For a circuit of length $l$ and a gate set of size $G$, the search space can grow as $G^{l}$, making an exhaustive search for an optimal circuit intractable.

This challenge is profoundly exacerbated by the lack of a smooth metric to guide the search. The standard measure of success, gate fidelity, is a global property of the entire matrix and provides a notoriously sparse signal for iterative optimization. For example, a circuit that is only a single Pauli-$X$ gate away from the target identity matrix has a fidelity of exactly zero \cite{b6}. This cliff-like reward landscape provides no gradient or sense of direction, making it extremely difficult for search algorithms to know if they are close to a solution. This fundamental issue necessitates methods that can navigate a vast search space without relying on a well-behaved cost function.

Recent advancements have leveraged powerful machine learning frameworks to navigate this space, but they face a critical trade-off between \textit{inference speed} and \textit{solution diversity}. Deep reinforcement learning (RL) approaches, such as AlphaZero, have demonstrated remarkable strength in finding highly optimized circuits \cite{b4}. However, these methods are notoriously sample-inefficient, require exceptionally long training times, and are designed to converge to a single optimal policy \cite{b7}. This focus on one solution is a significant drawback in quantum compilation, where a \textit{diverse} set of valid, high-fidelity circuits is often required to find one that best matches a specific, restrictive hardware topology. Conversely, generative diffusion models like genQC \cite{b5} can produce a diverse set of circuits, but they suffer from slow inference \cite{b8, b9}. Their iterative refinement process requires a large number of sampling steps and evaluations to find a single correct circuit, making them impractical for on-the-fly compilation.

To address this gap, we propose a novel framework that is both \textbf{fast at inference} and \textbf{capable of generating a diverse set of high-quality solutions}. Our approach pairs a Generative Flow Network (GFlowNet) \cite{b10, b11} with a powerful, Transformer-based policy network \cite{b12}. We chose the following components specifically for Unitary Synthesis:

\begin{enumerate}
    \item \textbf{GFlowNets:} This framework is perfectly suited for our goal. Originating from fields like drug discovery, GFlowNets are designed to learn a policy that samples candidates (e.g., molecules, or in our case, circuits) with probability proportional to their final rewards. Instead of collapsing to a single best solution as in RL, GFlowNets naturally learn to generate a diverse set of high-reward candidates, a key requirement for practical synthesis.

    \item \textbf{Transformer-based policy:} The task of synthesis requires understanding the \textit{global} structure of the target unitary, where distant elements in the matrix are non-locally correlated. Transformers' attention mechanism \cite{b12} is particularly capable of capturing these relationships, allowing them to read the design of the entire matrix and compress its high-dimensional state into a dense latent representation. This latent vector then allows a final policy network to efficiently learn and output a distribution over optimal next actions.
\end{enumerate}
A core methodological contribution of our work lies in how we formulate the problem. Rather than defining a reward function $R(U)$ that changes for every new target matrix $U$, we reframe the entire synthesis task as a path-finding problem toward a \textit{single, universal goal}. We define the agent's state as the \emph{unitary residual} $s_t = U V_t^\dagger$, where $U$ is the target and $V_t$ is the circuit built so far. The task is thus transformed into finding a sequence of gates that navigates from any \textit{initial state} $s_0 = U$ (for an empty circuit $V_0 = I$) to the \textit{fixed, universal terminal state} $s_\mathrm{f} = I$ (the identity matrix). This unified formulation makes our terminal reward function, $R(s_\mathrm{f})$, completely independent of the specific $U$ being synthesized. The GFlowNet learns a single general policy to reach the identity matrix, while the Transformer learns to map any given state $s_t$ to the best action to get there.

Our framework, QFlowNet, learns to solve this problem efficiently using only the sparse terminal reward, eliminating the need for complex reward shaping or pre-trained models. Our results demonstrate that our agent achieves a 99.7\% success rate on a 3-qubit benchmark of random unitaries across circuit lengths from 1 to 12. Crucially, it provides superior inference efficiency over diffusion models and discovers a diverse set of compact circuits that are often shorter than those found by standard compilers. We establish that this GFlowNet-Transformer hybrid is an effective, efficient, and versatile paradigm for unitary synthesis.

\section{Preliminaries}

\subsection{Unitary Synthesis and Its Challenges}
\label{sec:prelim_synthesis}
Unitary Synthesis is the process of decomposing a given $d \times d$ unitary matrix $U$ (where $d=2^n$ for $n$ qubits) into a sequence of gates from a predefined basis gate set $G$ \cite{b1}.

The challenge is twofold. First, the problem is \textbf{combinatorial}, with a search space growing as $O(G^{l})$ for circuit length $l$ \cite{b13, b14}. Second, the reward landscape is \textbf{sparse}; fidelity provides no gradient, as a nearly-correct circuit can have zero fidelity \cite{b6, b15, b16}.

Early ML work, such as that by Moro et al. \cite{b7}, focused on single-qubit synthesis, decomposing unitaries into fine-grained rotations like $\pi/64$ gates. The problem becomes exponentially harder in multi-qubit systems. The introduction of entangling gates, such as CNOT ($CX$), dramatically expands the search space and induces complex correlations \cite{b18, b19}.

A key challenge is the sheer size of the target space $U$. A model trained to synthesize any arbitrary unitary must also learn to generate decompositions for highly complex, random-looking unitaries that may have no practical algorithmic use. This ``useless unitary'' problem inflates the task difficulty. To make the problem tractable, generative models often restrict the domain. A common approach, used by genQC \cite{b5} and this work, is to limit the target unitaries to those reachable within a maximum gate length $L_\mathrm{max}$.

While our work is on exact synthesis, approximate synthesis is often tackled using variational circuits with parameterized gates (e.g., $R_X(\theta)$) \cite{b20, b21, b22, b23}. It is also possible to use GFlowNets with continuous action spaces to find these parameters. Finally, traditional RL methods are often formulated to use a single gate set, which is a drawback when a diverse set of valid circuits is needed. The \textit{Clifford+$T$ gate set} is recognized as universal \cite{b19, b24}, but its $T$ gate is often costly \cite{b25}. Our framework is flexible and can be trained on any custom discrete gate set \cite{b5, b26}.

\subsection{GFlowNets}
\label{sec:gflownet_prelim}

GFlowNets are a class of probabilistic models designed to learn a stochastic policy for generating compositional objects $x$ with probability $P(x)$ proportional to a given reward function, $R(x)$ \cite{b10, b11}. This objective, $P(x) \propto R(x)$, distinguishes GFlowNets from two other major paradigms.

First, unlike standard RL methods, which are trained to find a single optimal policy $\pi^*$ that maximizes expected return (i.e., finding a single $\arg\max_x R(x)$) \cite{b27}, GFlowNets are designed to learn a policy that samples a \textit{diverse} set of high-reward candidates \cite{b30}. Second, unlike many other generative models (e.g., VAEs, GANs, or diffusion models \cite{b8, b26, b28, b29}), which learn to model a fixed data distribution $p(x)$, GFlowNets are explicitly trained to match a given, non-negative reward function $R(x)$. This capability is particularly useful in scientific discovery, such as drug \cite{b31} or crystal \cite{b32} generation, where the goal is to generate many novel, high-scoring candidates.

GFlowNets model the generation process as a ``flow'' through a directed acyclic graph (DAG). The process starts at a single source node $s_0$ (e.g., an empty circuit) and moves through intermediate states by applying actions. The process terminates when a terminal node $s_\mathrm{f}$ (a complete object) is reached, at which point a sparse terminal reward $R(s_\mathrm{f})$ is received.

The core of GFlowNet is to learn a forward policy $P_\mathrm{F}(s' \mid s)$. The learning objective is to ensure ``flow consistency''. A common and effective method to achieve this is the \textbf{Trajectory Balance (TB)} objective \cite{b33}. The TB loss forces the probability of a forward-sampling trajectory $\tau = (s_0 \to \cdots \to s_\mathrm{f})$ to match the target reward, normalized by a learned \emph{partition function} $Z$ (the sum of all rewards) \cite{b11, b33}. The TB loss is defined as

\begin{equation}
    \mathcal{L}_{\text{TB}}(\tau; \theta) = \left( \log Z_\theta + \sum_{s \to s' \in \tau} \log P_\mathrm{F}(s' \mid s; \theta) - \log R(s_\mathrm{f}) \right)^2 \label{eq1}.
\end{equation}

By minimizing this loss, the reward signal $R(s_\mathrm{f})$ is backpropagated through all actions in the trajectory, solving the sparse-reward credit assignment problem \cite{b17, b33, b34}.

\section{Methods}

\begin{figure}[h]
\centering
     \includegraphics[page=1, width=1\linewidth]{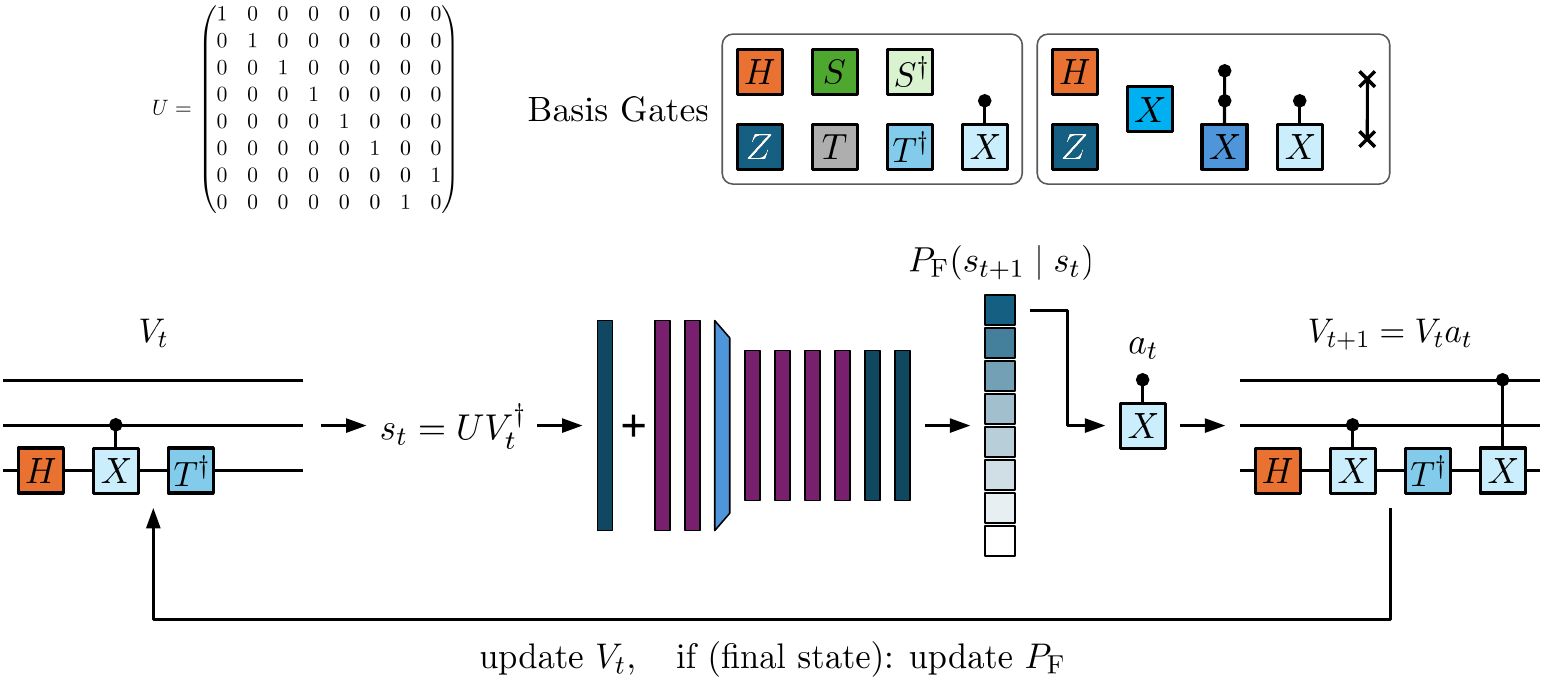}
\caption{\textbf{Schematic of the QFlowNet framework with Transformer policy.} At each step $t$, the agent observes the unitary residual $s_t =UV_t^\dagger$. This state is processed by the policy network (a Transformer-based unitary model, shown as the block architecture), which outputs a probability distribution $P_\mathrm{F}(s_{t+1} \mid s_t)$ over all possible next actions (gates). An action $a_t$ is sampled to determine the next state $s_{t+1}$, updating the synthesized circuit ($V_{t+1} = V_t a_t$). The process repeats until a final state is reached. A terminal reward is then calculated and used to update the policy network $P_\mathrm{F}$, completing the learning loop.}

\label{fig1}
\end{figure}

This section details our proposed QFlowNet framework, which covers the problem formulation, model architecture, and reward function design.
Fig.~\ref{fig1} provides a high-level schematic of the QFlowNet learning loop.

\subsection{Problem Formulation as a Markov Decision Process}

We formulate the task of exact unitary synthesis as a sequential decision-making problem, modeled as a Markov Decision Process (MDP).

\begin{itemize}
\item \textbf{State space $\mathcal{S}$}: A state $s_t\in \mathcal{S}$ is defined by the unitary residual $s_t = U V_t^\dagger$ where $U$ is the constant target unitary, and $V_t$ is the unitary of the circuit constructed by the agent after $t$ steps. The initial state is $s_0 =U$, corresponding to an empty circuit ($V_0 =I$). The terminal state is the identity matrix $I$.

\item \textbf{Action space $\mathcal{A}$}: The action space is discrete and consists of applying a single gate from a predefined basis gate set $G$. As discussed in Section \ref{sec:prelim_synthesis}, we use two different universal sets: $G_1=\{H, Z, S, S^\dagger, T, T^\dagger, \text{CNOT}\}$ and $G_2=\{H,X,Z, \text{CNOT}, \text{CCNOT}, \text{SWAP}\}$.

\item \textbf{Trajectories $\tau$}: An episode consists of generating a trajectory $\tau=(s_0 \to s_1 \to \cdots \to s_\mathrm{f})$ by sequentially applying actions. The episode terminates when the synthesized unitary $V_\mathrm{f}$ successfully matches the target $U$ or when a maximum sequence length $L_\mathrm{max}$ is reached.
\end{itemize}

\subsection{QFlowNet Framework for Circuit Synthesis}

We employ a GFlowNet to learn a policy for generating circuit trajectories with the TB objective as mentioned in \cite{b10, b11, b33}.

Here, the model learns the forward policy $P_\mathrm{F}(s_{t+1} \mid s_t;\theta)$ and the partition function $Z_\theta = \sum_\tau R(s_\mathrm{f})$. For simplicity, we assume a uniform backward policy. Hence, the term for $P_\mathrm{B}$ is omitted. This is a common simplification \cite{b17, b34}, and we found it sufficient for this problem space.

\begin{algorithm}
\caption{Trajectory Likelihood Maximization}\label{alg:gflownet}
\begin{algorithmic}[1] 

\STATE \textbf{Input:} Forward and backward parameters $\theta_\mathrm{F}^1, \theta_\mathrm{B}^1$, any GFlowNet loss function $\mathcal{L}_{\text{TB}}$, \textit{(optional)} experience replay buffer $\mathcal{B}$;

\FOR{$t = 1$ to $N_{\text{iters}}$}
    \STATE Sample a batch of trajectories $\left\{\tau_k^{(t)}\right\}_{k=1}^K$ from the forward policy $P_\mathrm{F}\!\left(\cdot \mid \cdot, \theta_\mathrm{F}^t\right)$;
    \STATE \textit{(optional)} Update $\mathcal{B}$ with $\left\{\tau_k^{(t)}\right\}_{k=1}^K$;
    \STATE Update $\theta_\mathrm{B}^{t+1} = \theta_\mathrm{B}^t - \gamma_t \cdot \frac{1}{K} \sum_{k=1}^K \nabla_{\theta_\mathrm{B}^t} \mathcal{L}_{\text{TB}}\!\left(\theta_\mathrm{B}^t; \tau_k^{(t)}\right)$;
    \STATE \textit{(optional)} Resample a batch of trajectories $\left\{\tau_k^{(t)}\right\}_{k=1}^K$ from $\mathcal{B}$;
    \STATE Update $\theta_\mathrm{F}^{t+1}$\\
    $= \theta_\mathrm{F}^t - \eta_t \cdot \frac{1}{K} \sum_{k=1}^K \nabla_{\theta_\mathrm{F}^t} \mathcal{L}_{\text{TB}}\!\left(\theta_\mathrm{F}^t; \tau_k^{(t)}, P_\mathrm{B}\!\left(\cdot \mid \cdot, \theta_\mathrm{B}^{t+1}\right)\right)$;
\ENDFOR 

\end{algorithmic}
\end{algorithm}

This forward policy $P_\mathrm{F}$ is parameterized by our \texttt{QFlowNet} model. This model is composed of two main parts: a \texttt{Unitary\_encoder} that processes the state, and a \texttt{policy\_head} that outputs action logits.

The \texttt{Unitary\_encoder} is a hybrid CNN-Transformer architecture designed to process the $2\times d\times d$ state tensor. It uses an initial convolutional layer to project the input into a high-dimensional feature space, followed by a series of Transformer-based self-attention blocks to capture non-local relationships across the matrix. This encoder block outputs a condensed sequence of embeddings.

The \texttt{policy\_head} is a simple Multi-Layer Perceptron (MLP) that maps the final pooled state embedding from the encoder to the logits over the discrete action space $\mathcal{A}$.

This design effectively separates the two most critical tasks: the Transformer-encoder is responsible for \textit{state understanding} by compressing the complex unitary into a meaningful latent vector, while the MLP head is responsible for \textit{decision making} by translating that vector into a policy. A detailed specification of each layer, including kernel sizes, attention heads, and block depths, is provided in Appendix~\ref{secA1}.

\subsection{Reward Function Design}

\begin{figure}[h]
\centering

 \begin{subfigure}[b]{1\linewidth}
     \centering
     \includegraphics[page=2, width=\textwidth]{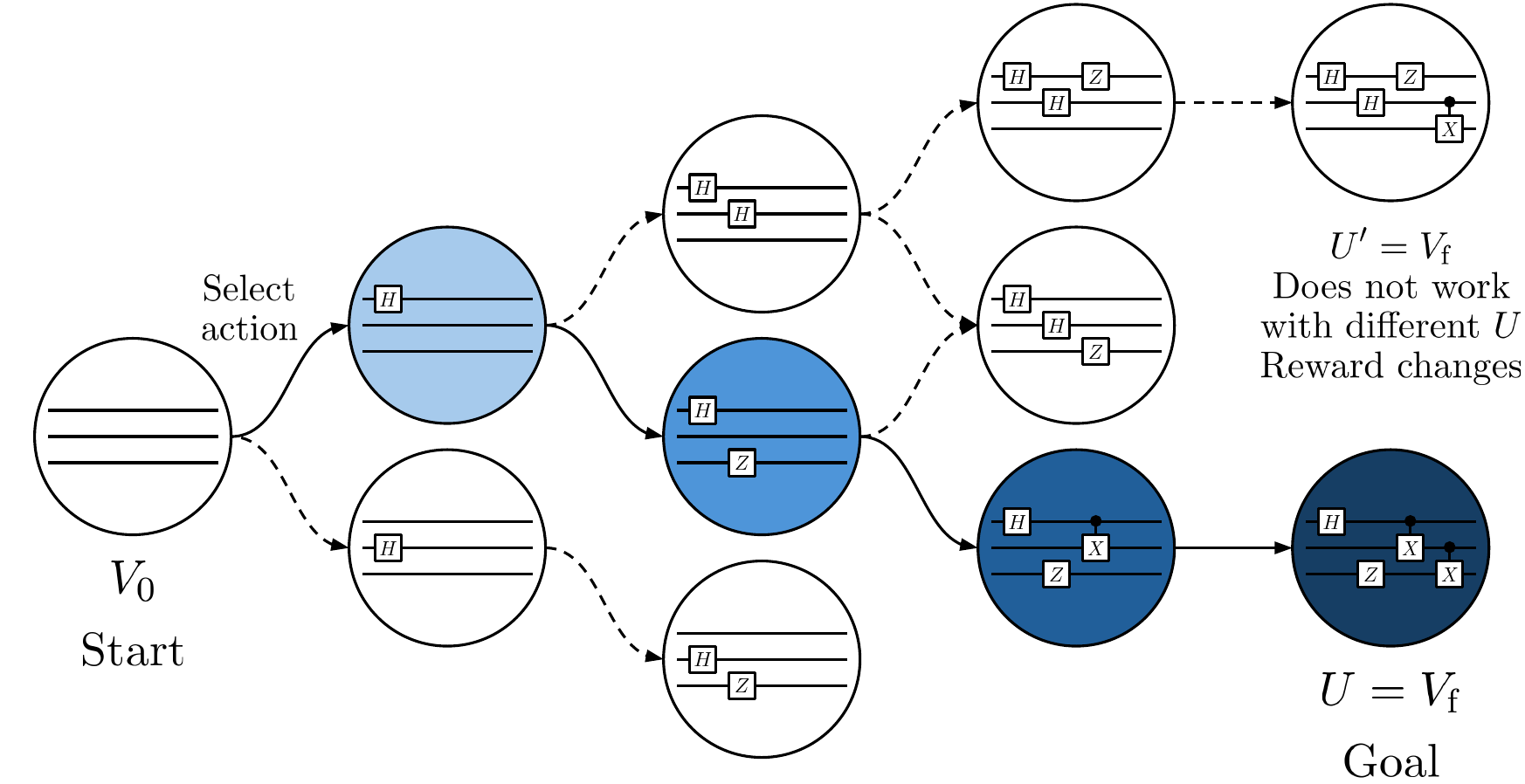}
     \caption{}
 \end{subfigure}
 \begin{subfigure}[b]{1\linewidth}
     \centering
     \includegraphics[page=3, width=\textwidth]{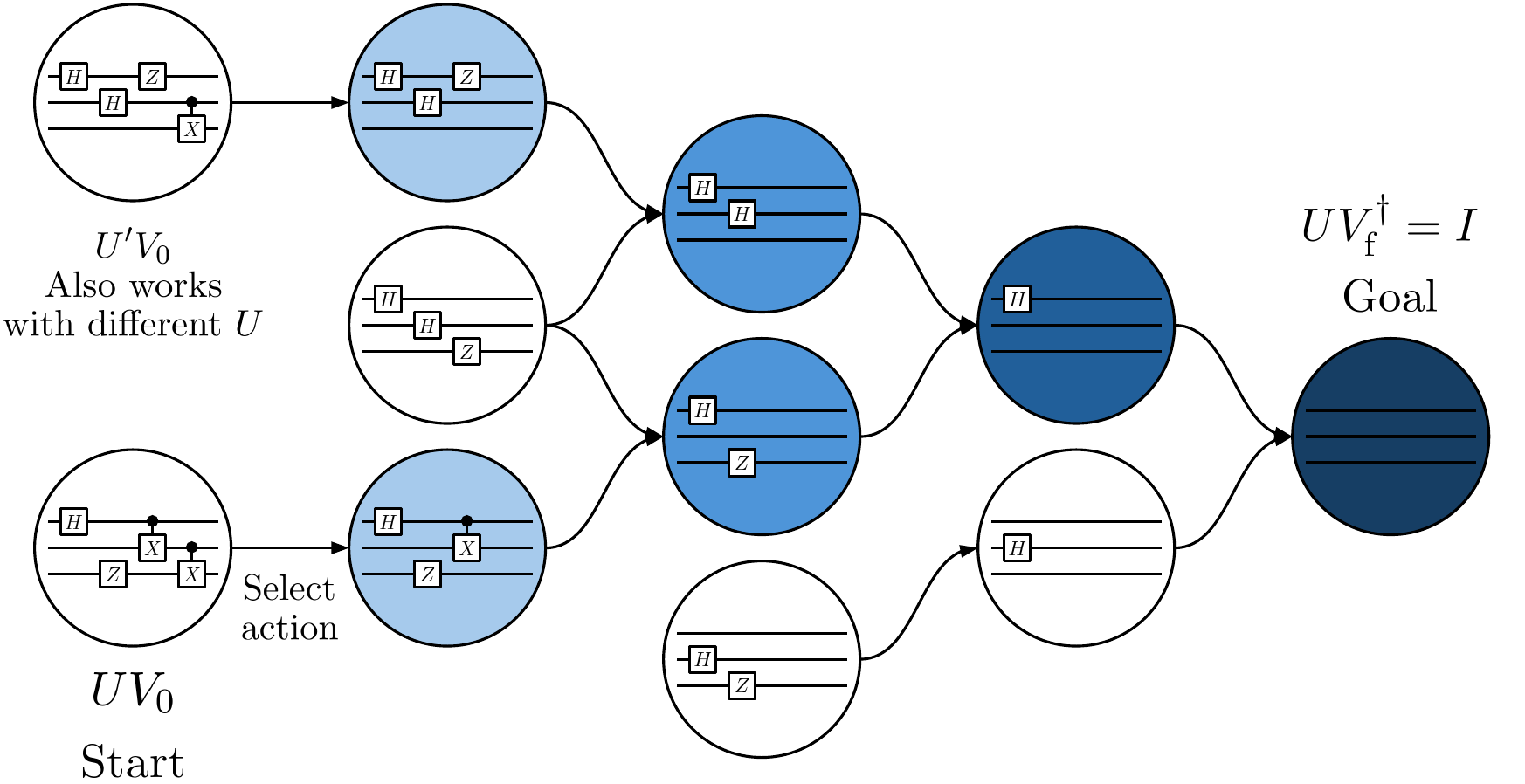}
     \caption{}
 \end{subfigure}

\caption{\textbf{Conceptual comparison of synthesis problem formulations.}
\textbf{(a)} A standard formulation where the agent starts from a fixed empty circuit ($V_0$) and attempts to reach a target-dependent goal state ($U$). This approach requires the reward function itself (e.g., $\tr\big(U'V_\mathrm{f}^\dagger\big)$) to be redefined for every new target $U'$, precluding policy reuse.
\textbf{(b)} Our proposed QFlowNet formulation. The problem is reframed by defining the state as the ``unitary residual'' ($s_t = U V_t^\dagger$). The agent now starts from a target-dependent \textit{start state} ($s_0 = U$) and navigates to a \textit{fixed, universal goal state} ($s_\mathrm{f} = I$). This design makes the reward function (fidelity to $I$) universal, allowing a single, general policy to be trained and applied to any target unitary matrix.}
\label{fig2}
\end{figure}

A key methodological contribution of our framework is the formulation of the reward function, which is intrinsically linked to our state definition. As shown in Fig.~\ref{fig2}, the state $s_t = U V_t^\dagger$ transforms the synthesis problem into a path-finding task. The agent's goal is to find a path (a sequence of gates) from a problem-specific \textit{start state} ($s_0 = U$) to a \textit{universal, fixed target state} ($s_\mathrm{f} = I$).

This formulation allows for a reward function that is completely independent of the specific target unitary $U$. Unlike other GFlowNet applications, where the reward function $R(x)$ must be redefined for each new target, our reward function is universal. The agent is trained once on the fixed objective of reaching the identity matrix $I$. This single, general policy can then be applied to synthesize \textit{any} target unitary $U$ simply by changing the initial state $s_0$ given to the agent.

The design of the reward function is also critical. Our framework relies exclusively on a sparse but effective terminal reward that directly reflects this universal objective. Fidelity is defined as $F(U, V_\mathrm{f}) = \frac{1}{d} |\tr(U^\dagger V_\mathrm{f})|$. For a trajectory $\tau$ that terminates in a state $s_\mathrm{f}$, the reward $R(\tau)$ is defined as:

\begin{equation}
R(\tau) = 
\begin{cases}
    R_\mathrm{success} & \text{if }F(U, V_\mathrm{f}) > 0.999 \\
    \epsilon & \text{otherwise}
\end{cases},
\label{eq2}
\end{equation}
where $R_\mathrm{success}$ is a large positive constant ($=10^{2}$) and $\epsilon$ is a small positive constant ($=10^{-4}$) to ensure all paths have nonzero flow. This binary, outcome-based reward structure is maximally sparse. The success of our model demonstrates that the QFlowNet objective is capable of effective credit assignment based on this simple, delayed signal.

\subsection{Training and Evaluation Protocol}
\label{secTE}

\begin{itemize}
\item \textbf{Training}: The agent was trained for 100,000 steps with a batch size of 2048 in the 3-qubit experiment. Due to memory limitations, the 4-qubit experiment and 5-qubit experiment were trained with 50,000 steps with a batch size of 1024 and 50,000 steps with a batch size of 128. For each trajectory in the batch, a new target unitary $U$ was generated by creating a random circuit with a depth sampled uniformly from $\{1, \dots , 12\}$. We used the Adam optimizer with a learning rate of $1\times10^{-4}$.

\item \textbf{Evaluation}: The trained model was evaluated on a test set. For each circuit depth $L \in \{1, \dots , 12\}$, we generated 100 distinct random circuits of that depth. To create a challenging and realistic baseline, each of these circuits was then optimized using the \texttt{transpile} method of Qiskit \cite{b35} (\texttt{optimization\_level=1}). The resulting unitaries formed our test set. For each target unitary, we sampled up to 1024 candidate circuits. A synthesis was marked as successful if any candidate achieved fidelity $> 0.999$.
\end{itemize}

\section{Results}

To evaluate our QFlowNet framework, we conducted a series of experiments focusing on the exact synthesis of 3- to 5-qubit unitaries. Our evaluation is designed to answer three central questions: (1) How effectively can our agent synthesize complex, randomly generated unitaries? (2) How does its inference efficiency compare to other generative models? (3) Does the agent discover compact, optimized circuit implementations, and can it produce a diverse set of valid solutions?

\subsection{Synthesis Accuracy and Scalability}\label{AA}

\begin{figure}[H]
    \centering
    \begin{subfigure}[b]{0.49\linewidth}
        \centering
        \includegraphics[width=\textwidth]{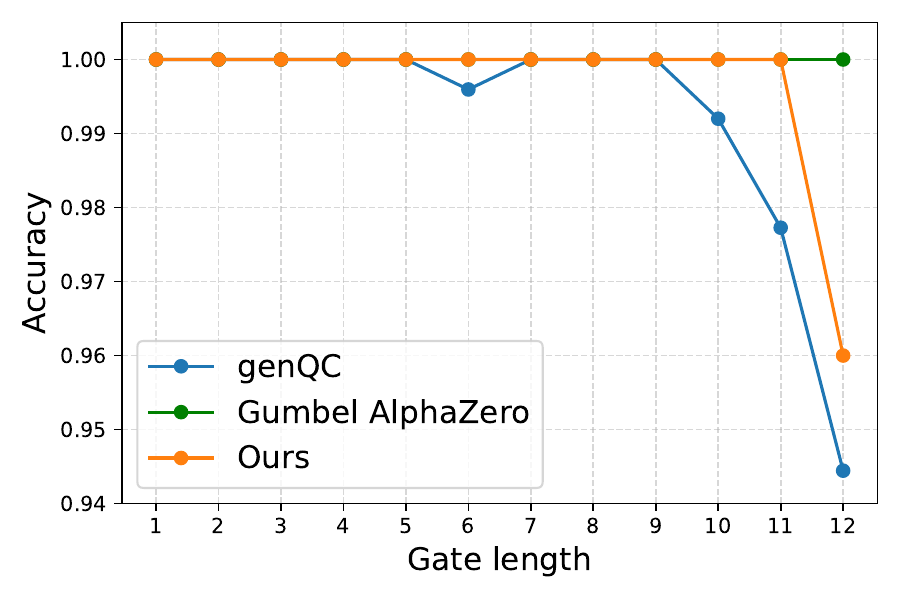}
        \caption{}
        \label{fig:3a}
    \end{subfigure}
    \hfill
    \begin{subfigure}[b]{0.49\linewidth}
        \centering
        \includegraphics[width=\textwidth]{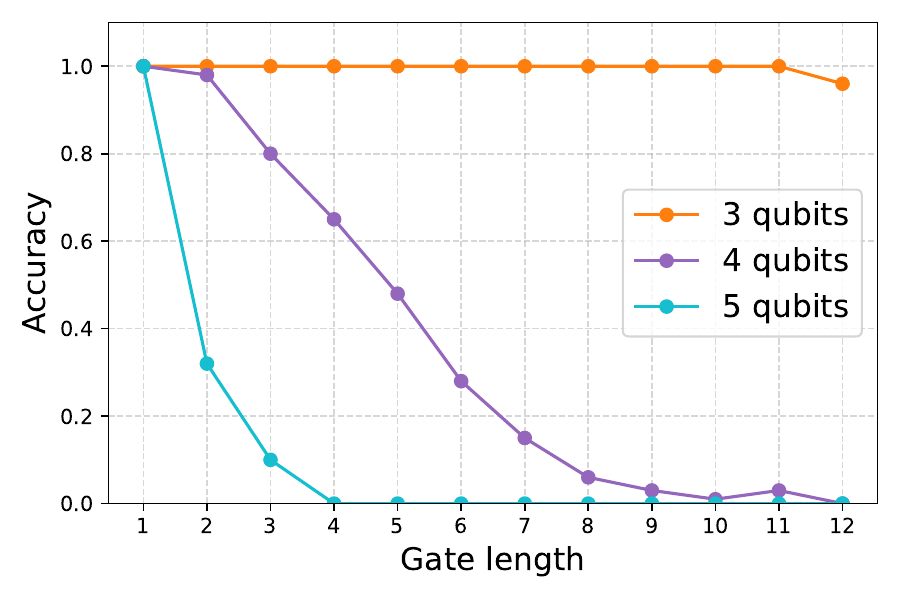}
        \caption{}
        \label{fig:3b}
    \end{subfigure}
    \caption{\textbf{Synthesis accuracy as a function of circuit complexity.} \textbf{(a)} The plot shows a comparison of the success rates on 3-qubit unitaries. \textbf{(b)} The plot compares the multi-qubit success rates of QFlowNet, highlighting the scalability challenge.}
    \label{fig3}
\end{figure}

We assessed our algorithm's ability to synthesize complex, randomly generated unitaries for 3-, 4-, and 5-qubit systems and compared the 3-qubit accuracy with the strong baseline generative model genQC \cite{b5} and the RL-based model Gumbel AlphaZero \cite{b25}. The test set was constructed as described in Section~\ref{secTE}. The agent was permitted to sample up to 1024 candidate circuits, with a synthesis considered successful if any candidate achieved the target fidelity.

The results presented in Fig.~\ref{fig3} indicate that our QFlowNet can effectively learn a robust synthesis policy from a sparse reward signal. As shown in Fig.~\ref{fig:3a}, the 3-qubit model demonstrates a remarkable ability to synthesize complex unitaries, achieving a success rate of 99.7\% across the entire benchmark. Performance remains high for circuits with optimal depths up to 10, with a success rate of 96\% on the most challenging depth-12 instances.


The multi-qubit results, detailed in Fig.~\ref{fig:3b}, underscore the challenge of scalability. While the 4-qubit agent achieves 100\% accuracy for simple circuits, its performance rapidly declines for more complex unitaries, dropping to 48\% by length 5 and to under 10\% for lengths 8 and greater. This performance drop is a clear illustration of the $O(4^{n})$ complexity bottleneck, as will be discussed in Section~\ref{sec:discussion}, which remains the primary hurdle for scaling this method.

\subsection{Inference Efficiency}

\begin{figure}
    \centering
    \begin{subfigure}[b]{0.49\linewidth}
        \centering
        \includegraphics[width=\textwidth]{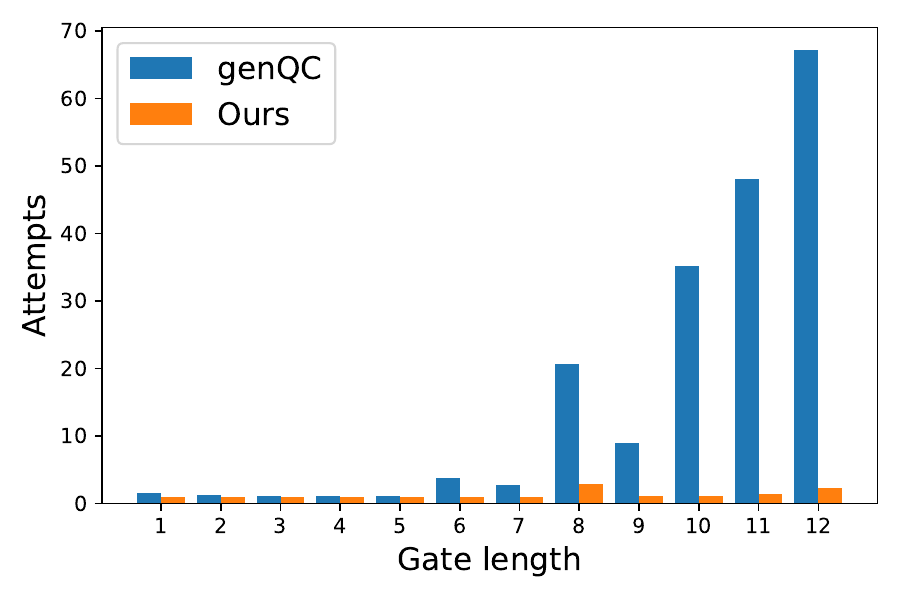}
        \caption{}
        \label{fig:4a}
    \end{subfigure}
    \hfill
    \begin{subfigure}[b]{0.49\linewidth}
        \centering
        \includegraphics[width=\textwidth]{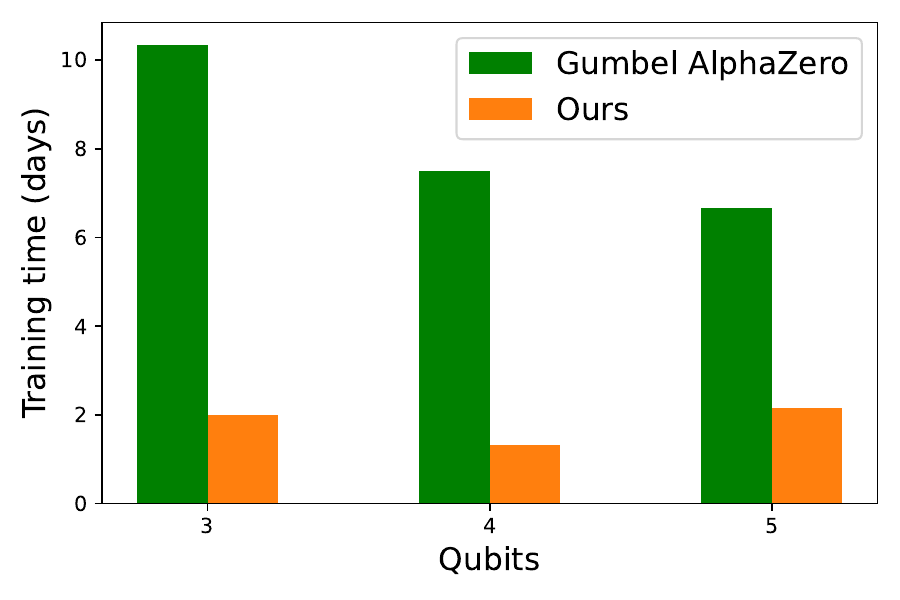}
        \caption{}
        \label{fig:4b}
    \end{subfigure}
    \caption{\textbf{Efficiency comparison of QFlowNet against other ML methods.} \textbf{(a)} Inference efficiency comparison. Our QFlowNet agent (orange) consistently requires only a small number of attempts, while the genQC diffusion model \cite{b5} (blue) requires an exponentially increasing number of samples. \textbf{(b)} Training time comparison. Total training time in days for our QFlowNet versus the Gumbel AlphaZero-based model \cite{b25}. Our agent trains in 1--2 days, while the Gumbel AlphaZero approach requires 6.5--10 days.}
    \label{fig4}
\end{figure}

Beyond accuracy, we evaluated the practical inference efficiency of our framework, comparing the number of sampling attempts required by our agent against those required by the genQC diffusion model \cite{b5}. As illustrated in Fig.~\ref{fig:4a}, this comparison reveals a critical advantage of our approach.

For circuits of length 8 or greater, the sampling requirement for genQC increases significantly, culminating in an average of nearly 70 attempts for length-12 unitaries. In contrast, our QFlowNet agent exhibits remarkable stability, requiring only 1--2 attempts on average across the entire range of complexities.


This efficiency extends directly to training time. As shown in Fig.~\ref{fig:4b}, our QFlowNet-based framework is significantly more computationally feasible. For the 3-qubit synthesis task, our model converges in 2 days, whereas the Gumbel AlphaZero approach \cite{b25} requires over 10 days. This advantage holds across scales. The results empirically validate our claim that the QFlowNet paradigm avoids the sample-inefficiency and the notoriously long training time \cite{b4, b25} associated with model-based RL.

\subsection{Discovery of Compact and Diverse Circuit Solutions}

\begin{figure}
    \centering
    \begin{subfigure}[b]{0.49\linewidth}
        \centering
        \includegraphics[width=\textwidth]{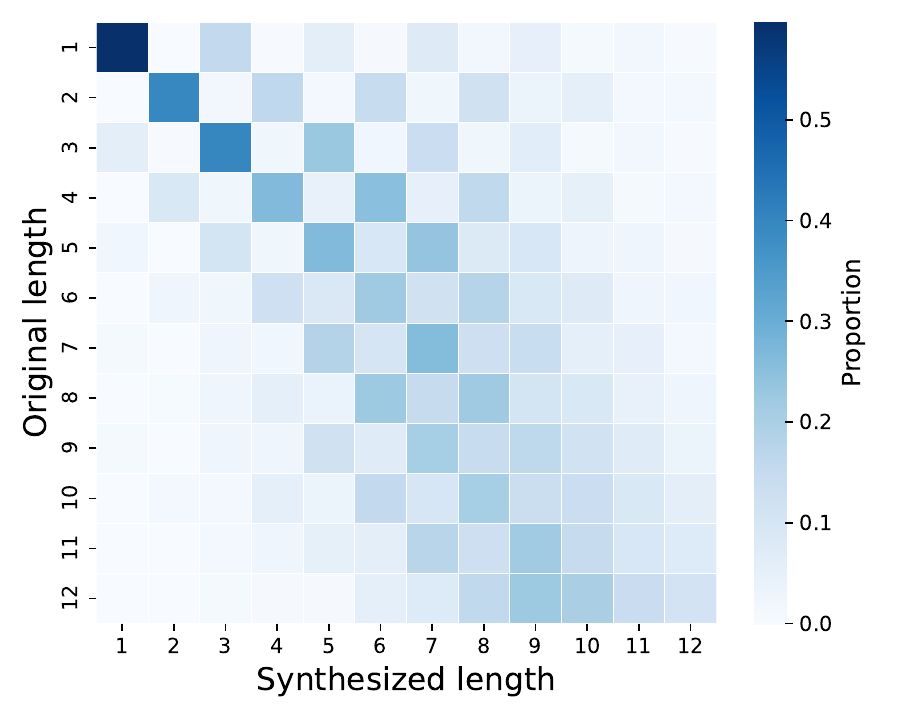}
        \caption{}
        \label{fig:5a}
    \end{subfigure}
    \hfill
    \begin{subfigure}[b]{0.49\linewidth}
        \centering
        \includegraphics[width=\textwidth]{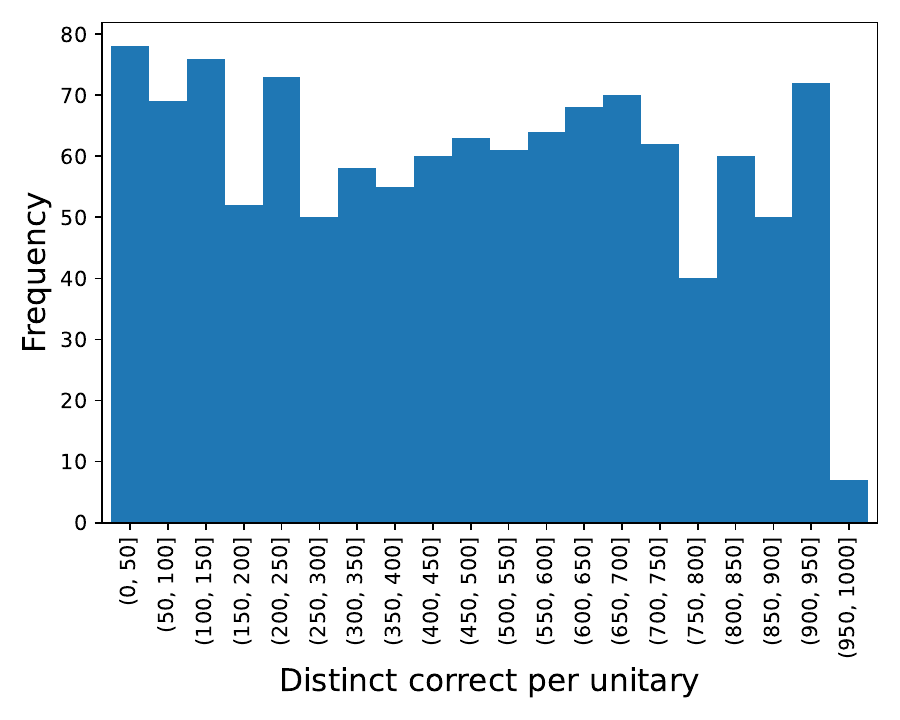}
        \caption{}
        \label{fig:5b}
    \end{subfigure}
    \caption{\textbf{Analysis of synthesized circuit compactness and diversity.} \textbf{(a)} The normalized confusion matrix displays the distribution of synthesized circuit lengths (x-axis) versus the Qiskit-optimized target lengths (y-axis). The most notable results are the strong diagonal (length-optimal) and the cells below it (more compact than the baseline). \textbf{(b)} The histogram shows the number of distinct, correct circuits discovered per target unitary (based on 1024 samples).}
    \label{fig5}
\end{figure}

Finally, we analyzed the quality of the circuits discovered by our agent, focusing on both compactness and generative diversity. Fig.~\ref{fig5} presents these two aspects.


First, we analyzed compactness, as shown in Fig.~\ref{fig:5a}. We compared our synthesized length against the Qiskit-optimized baseline. The strong diagonal concentration indicates that the model most frequently generates circuits with the correct optimal length. Most notably, a significant number of syntheses result in circuits that are shorter than the Qiskit-optimized baseline (cells below the diagonal).

Second, we analyzed the generative diversity, as shown in Fig.~\ref{fig:5b}. A key feature of a generative approach is the ability to sample multiple distinct solutions. As shown in the histogram, the agent frequently discovers dozens or even hundreds of distinct, valid decompositions for a single target. This generative capability is a significant advantage over deterministic search methods and is a property shared by other generative models \cite{b5, b26}.

\section{Discussion}\label{sec:discussion}

In this work, we introduce QFlowNet, a novel framework for exact unitary synthesis. We observed that the vast majority of syntheses were successful on the first attempt, indicating that QFlowNet learns a policy that is sharply focused on valid and compact solutions. This challenges the notion that complex reward shaping \cite{b36, b37} is necessary. We employ a Transformer architecture because its self-attention mechanism \cite{b12} is well-suited to capture the long-range dependencies within a unitary matrix. This concept is validated by other works \cite{b4, b5} that also leverage a Transformer-based `Unitary encoder'.

Our framework offers notable advantages over existing ML methods. Compared to deep RL approaches like Gumbel AlphaZero \cite{b25, b27}, our QFlowNet method is more computationally feasible. GFlowNets are designed to learn from terminal rewards, which they efficiently distribute via objectives like TB \cite{b33}. Consequently, we avoid the difficult process of designing intermediate reward functions.

Furthermore, the GFlowNet framework is inherently designed to generate a diverse set of high-reward candidates \cite{b10, b30, b31}. As demonstrated in the results in Fig.~\ref{fig:5b}, the agent discovers hundreds of distinct, valid circuits. This allows a user to sample a variety of circuits and select the optimal one based on secondary metrics, such as CNOT count or hardware connectivity.
Compared with other generative models, our approach demonstrates superior inference efficiency. As empirically illustrated in Fig.~\ref{fig:4a}, diffusion models such as genQC \cite{b5, b8, b9} require an exponentially increasing number of samples to synthesize complex unitaries. In contrast, our QFlowNet learns a policy that is sharply peaked around valid solutions, enabling it to generate a correct circuit in a single pass for the vast majority of cases.

The success of our approach suggests broader implications for the automated design of quantum circuits. It validates the idea that complex, combinatorial search problems in quantum computing can be effectively solved by pairing a powerful generative model such as GFlowNet with an expressive architecture like the Transformer. This combination is adept at learning from sparse, outcome-based rewards, converting an intractable search problem into a guided, sequential decision-making process without manual intervention. This strategy could be extended to other challenging problems in quantum control and compilation, such as quantum architecture search \cite{b20, b23, b38} or pulse-level control optimization \cite{b39}.

Despite these promising results, our work has several limitations that provide avenues for future research. The primary challenge remains scalability. 
While we have successfully trained models for 3-, 4-, and 5-qubit synthesis, our method in its current form relies on the full $2 \times 2^n \times 2^n$ tensor as input, which becomes computationally impractical for larger systems. The sharp decline in 4-qubit performance shown in Fig.~\ref{fig:3b} is a direct consequence of this $O(4^{n})$ complexity, which remains a fundamental hurdle. However, these results are attributed to memory limitations. While the batch size and training steps are directly related to the number of states the agent visits, increasing the computing power and memory allocation appears to be a solution to the problem of scalability. Alternatively, future investigations should explore more efficient representations, such as decomposing the problem \cite{b30, b40}, or conditioning strategies that do not require the entire matrix. Second, our study was confined to a discrete gate set. A natural extension is to incorporate continuous, parameterized gates. Finally, while our Transformer-based policy network proved effective, exploring different architectural variations or conditioning methods could further enhance the agent's performance. Adapting the framework from exact to approximate synthesis also remains a key direction for future work.

\section*{Acknowledgment}

This work is in part supported by the National Research Foundation of Korea (NRF, RS-2024-00451435 (20\%), RS-2024-00413957 (20\%)), Institute of Information \& communications Technology Planning \& Evaluation (IITP, RS-2025-02305453 (15\%), RS-2025-02273157 (15\%), RS-2025-25442149 (15\%) RS-2021-II211343 (15\%)) grant funded by the Ministry of Science and ICT (MSIT), Institute of New Media and Communications (INMAC), and the BK21 FOUR program of the Education, Artificial Intelligence Graduate School Program (Seoul National University), and Research Program for Future ICT Pioneers, Seoul National University, and Korea Institute for Advancement of Technology (KIAT) grant funded by the Korea Government (Ministry of Education) (P0025681-G02P22450002201-10054408, Semiconductor-Specialized University) in 2026.


\vspace{12pt}
\renewcommand{\appendixname}{Appendix A}
\appendix
\section{Model Architecture}
\label{secA1}

The \texttt{QFlowNet} policy network is composed of a \texttt{Unitary\_encoder} that extracts features from the state and a \texttt{policy\_head} that outputs action logits.

The \texttt{Unitary\_encoder} utilizes a hybrid CNN-Transformer architecture. It takes a $2\times d\times d$ tensor representation of the $n$-qubit unitary matrix (split into real and imaginary channels) as input. The input is first projected into a feature space of dimension $D_{1}=64$ via a $1\times1$ convolution. A \texttt{PositionalEncoding2D} \cite{b15} is added to embed spatial information, followed by the first \texttt{SpatialTransformerSelfAttn} block. This block has a depth of 4 and uses 8 attention heads, maintaining the $d \times d$ spatial resolution.

Subsequently, the features are down-sampled by a factor of 2 (resulting in a resolution of $d/2 \times d/2$) using a \texttt{DownBlock2D}, which simultaneously doubles the feature dimension to $D_{2}=128$. A second \texttt{SpatialTransformerSelfAttn} block (depth 4, 8 heads) processes these down-sampled features. A final $1\times1$ convolution projects the features onto the final embedding size ($D_\mathrm{emb}=256$).

To generate the policy, the encoder output sequence is aggregated via global mean pooling (\texttt{z\_seq.mean(dim=1)}) to produce a single state vector $z \in \mathbb{R}^{256}$. This vector is passed to the \texttt{policy\_head}, which is an MLP consisting of a linear layer ($256 \to 128$), a ReLU activation, and a final linear layer mapping to the logits over the discrete actions in $\mathcal{A}$.

\begin{figure}
    \centering
    \includegraphics[width=1\linewidth]{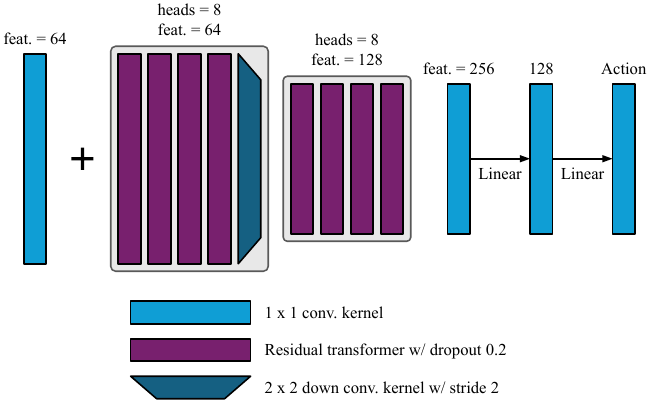}
    \caption{\textbf{Schematic overview of the QFlowNet architecture.} The model processes the input unitary matrix through a hybrid CNN-Transformer encoder, consisting of two stages of spatial Transformer blocks (depth 4, 8 heads) separated by a down-sampling layer. The resulting 256-dimensional state embedding is aggregated and passed to the MLP (hidden dimension 128) to predict action logits.}
\label{fig6}
\end{figure}





\begin{thebibliography}{50}
\bibitem{b1} M. A. Nielsen and I. L. Chuang, Quantum Computation and Quantum Information, Cambridge: Cambridge University Press, 2010.
\bibitem{b2} W. Zhang, J. Liu, Z. Zhou, and S. Yang, ``Approximate quantum circuit synthesis for diagonal unitary,'' arXiv preprint arXiv:2402.01869, 2024.
\bibitem{b3} C. M. Dawson and M. A. Nielsen, ``The Solovay-Kitaev algorithm,'' Quantum Inf. Comput., vol. 6, no. 1, pp. 81-95, 2006.
\bibitem{b4} F.J.R. Ruiz, T. Laakkonen, J. Bausch et al., ``Quantum circuit optimization with AlphaTensor,'' Nat Mach Intell 7, pp. 374–385, 2025. 
\bibitem{b5} F. Fürrutter, G. Muñoz-Gil, and H.J. Briegel, ``Quantum circuit synthesis with diffusion models,'' Nat Mach Intell 6, pp. 515–524, 2024.
\bibitem{b6} E. Magesan, J. M. Gambetta, and J. Emerson, ``Scalable and robust randomized benchmarking of quantum processes,'' Phys. Rev. Lett., vol. 106, no. 18, p. 180504, 2011.
\bibitem{b7} L. Moro, M. G. A. Paris, M. Restelli, and E. Prati, ``Quantum compiling by deep reinforcement learning,'' Commun. Phys., vol. 4, no. 1, pp. 1-8, 2021.
\bibitem{b8} J. Ho, A. Jain, and P. Abbeel, ``Denoising diffusion probabilistic models,'' Advances in Neural Information Processing Systems, vol. 33, pp. 6840-6851, 2020.
\bibitem{b9} Y. Song, J. Sohl-Dickstein, D. P. Kingma, A. K, E. G, and D. B, ``Score-based generative modeling through stochastic differential equations,'' arXiv preprint arXiv:2011.13456, 2020.
\bibitem{b10} E. Bengio, S. Lahlou, T. Deleu, M. Jain, B. Al-Omari, X. Zhang, and Y. Bengio, ``Flow Network based Generative Models for Non-Iterative Diverse Candidate Generation,'' Advances in Neural Information Processing Systems, vol. 34, pp. 26280-26291, 2021.
\bibitem{b11} Y. Bengio, S. Lahlou, T. Deleu, E. J. Hu, M. Tiwari, and E. Bengio, ``GFlowNet Foundations,'' J. Mach. Learn. Res., vol. 24, no. 210, pp. 1-55, 2023.
\bibitem{b12} A. Vaswani, N. Shazeer, N. Parmar, J. Uszkoreit, L. Jones, A. N. Gomez, {\L}. Kaiser, and I. Polosukhin, ``Attention is all you need,'' Advances in Neural Information Processing Systems, vol. 30, 2017.
\bibitem{b13} V. V. Shende, S. S. Bullock, and I. L. Markov, ``Synthesis of quantum-logic circuits,'' IEEE Transactions on Computer-Aided Design of Integrated Circuits and Systems, pp. 1000–1010, 2006.
\bibitem{b14} V. Kliuchnikov, D. Maslov, and M. Mosca, ``Fast and efficient exact synthesis of single-qubit unitaries generated by Clifford and T gates,'' Quantum Inf. Comput., vol. 13, no. 7-8, pp. 607--630, 2013.
\bibitem{b15} J. Schulman, F. Wolski, P. Dhariwal, A. Radford, and O. Klimov, ``Proximal policy optimization algorithms,'' arXiv preprint arXiv:1707.06347, 2017.
\bibitem{b16} A. Ecoffet, J. Huizinga, C. S, K. S, and W. O, ``First return, then explore,'' Nature, vol. 590, no. 7847, pp. 580-586, 2021.
\bibitem{b17} M. Pandey, G. Subbaraj, E. Bengio, ``GFlowNet Pretraining with Inexpensive Rewards,'' arXiv preprint arXiv:2409.09702, 2024.
\bibitem{b18} B. Kraus and J. I. Cirac, ``Optimal creation of entanglement using a two-qubit gate,'' Phys. Rev. A, vol. 63, no. 6, p. 062309, 2001.
\bibitem{b19} A. Barenco et al., ``Elementary gates for quantum computation,'' Phys. Rev. A, vol. 52, no. 5, p. 3457, 1995.
\bibitem{b20} M. Ostaszewski, L. M. Trenkwalder, W. Masarczyk, E. Scerri, and V. Dunjko, ``Reinforcement learning for optimization of variational quantum circuit architectures,'' Advances in neural information processing systems, vol. 34, pp. 18182-18194, 2021.
\bibitem{b21} C. H. Yang, H. Kim, J. Lee, D. K. Park, and J. Lee, ``Optimizing variational quantum circuits using reinforcement learning,'' Quantum Inf. Process., vol. 22, no. 1, p. 16, 2023.
\bibitem{b22} R. Tucci, ``An introduction to Cartan's KAK decomposition for QC programmers,'' arXiv preprint quant-ph/0507171, 2005.
\bibitem{b23} Y. J. Patel, K. S, S. S, and P. M, ``Curriculum reinforcement learning for quantum architecture search under hardware errors,'' arXiv preprint arXiv:2402.03500, 2024.
\bibitem{b24} N. J. Ross and P. Selinger, ``Optimal ancilla-free Clifford+T approximation of z-rotations,'' Quantum Inf. Comput., vol. 16, no. 11-12, pp. 901-956, 2016.
\bibitem{b25} S. Rietsch, J. Sbierski, T. B{\"a}ck, and V. Dunjko, ``Unitary synthesis of Clifford+T circuits with reinforcement learning,'' IEEE International Conference on Quantum Computing and Engineering (QCE), vol. 1, 2024.
\bibitem{b26} Wu, Jun, et al. ``Quantum circuit autoencoder.'' Physical Review A 109.3, 2024.
\bibitem{b27} D. Silver et al., ``A general reinforcement learning algorithm that masters chess, shogi, and Go through self-play,'' Science, vol. 362, no. 6419, pp. 1140-1144, 2018.
\bibitem{b28} Z. An, H. Zhou, S. L, H. L, K. H, and C. W, ``Quantum circuit synthesis for arbitrary logical operations using a quantum autoencoder,'' Opt. Express, vol. 29, no. 12, pp. 18485-18497, 2021.
\bibitem{b29} Z. Zhou, Y. Wu, J. W, Z. H, J. Z, J. L, and Y. D, ``Generative adversarial networks for quantum circuit synthesis,'' J. Quantum Inf. Sci., vol. 13, no. 2, pp. 83-96, 2023.
\bibitem{b30} Zhang, Dinghuai, et al. ``Let the flows tell: Solving graph combinatorial problems with gflownets,'' Advances in neural information processing systems, vol. 36, pp. 11952-11969, 2023.
\bibitem{b31} M. Jain, E. B, T. D, S. L, and Y. B, ``Biological sequence design with GFlowNets,'' International Conference on Machine Learning, 2022.
\bibitem{b32} Nguyen, Tri Minh, et al. ``Hierarchical gflownet for crystal structure generation,'' AI for Accelerated Materials Design-NeurIPS 2023 Workshop, 2023.
\bibitem{b33} N. Malkin, M. Jain, E. Bengio, C. Sun, and Y. Bengio, ``Trajectory balance: Improved credit assignment in GFlowNets,'' in Advances in Neural Information Processing Systems, vol. 35, pp. 15147-15159, 2022.
\bibitem{b34} Shen, W. Max, et al. ``Towards understanding and improving gflownet training,'' International conference on machine learning, 2023.
\bibitem{b35} Javadi-Abhari, Ali, et al. ``Quantum Computing with Qiskit.'' arXiv.org arxiv:2405.08810, 2024.
\bibitem{b36} T. F{\"o}sel, M. Y. Niu, F. Marquardt, and L. Li, ``Quantum circuit optimization with deep reinforcement learning,'' arXiv preprint arXiv:2103.07585, 2021.
\bibitem{b37} O. Lockwood, ``Optimizing quantum variational circuits with deep reinforcement learning.'' arXiv preprint arXiv:2109.03188, 2021.
\bibitem{b38} J. Landman, T. B{\"a}ck, and V. Dunjko, ``Deep symbolic regression for quantum-circuit synthesis,'' Phys. Rev. A, vol. 109, no. 4, pp. 042615, 2024.
\bibitem{b39} Smith et al., ``Programming physical quantum systems with pulse-level control,'' Frontiers in Physics, vol. 10, pp. 900099, 2022.
\bibitem{b40} Ni. Zhang, Zhiguang Cao. ``Hybrid-Balance GFlowNet for Solving Vehicle Routing Problems,'' arXiv preprint arXiv:2510.04792, 2025.
\end{thebibliography}
\end{document}